\documentstyle[aps,prl,epsfig,twocolumn]{revtex}

\def\be{\begin{equation}}
\def\ee{\end{equation}}
\def\bea{\begin{eqnarray}}
\def\eea{\end{eqnarray}}
\def\bma{\begin{mathletters}}
\def\ema{\end{mathletters}}

\newcommand{\eins}{\mbox{$1 \hspace{-1.0mm}  {\bf l}$}}
\def\C{\hbox{$\mit I$\kern-.7em$\mit C$}}

\begin{document}
\draft

\title{Optimal conversion of non--local unitary operations}

\author{W. D\"ur$^{1}$, G. Vidal$^2$ and J. I. Cirac$^{3}$}

\address
{$^1$Sektion Physik, Ludwig-Maximilians-Universit\"at M\"unchen, Theresienstr.\ 37,
D-80333 M\"unchen, Germany\\
$^2$Institut f\"ur Theoretische Physik, Universit\"at
  Innsbruck,A-6020 Innsbruck, Austria\\
$^3$Max--Planck Institut f\"ur Quantenoptik, Hans--Kopfermann Str. 1, D-85748 Garching, Germany}

\date{\today}

\maketitle

\begin{abstract}
We study when a non--local unitary operation acting on two $d$--level systems 
can probabilistically simulate another one when arbitrary local operations and 
classical communication are allowed. We provide necessary and sufficient 
conditions for the simulation to be possible. Probabilistic interconvertability is 
used to define an equivalence relation between gates. We show that this relation 
induces a finite number of classes, that we identify. In the case of two qubits, 
two classes of non--local operations exist. We choose the CNOT and SWAP as 
representatives of these classes. We show how the CNOT [SWAP] can be 
deterministically converted into any operation of its class. We also calculate 
the optimal probability of obtaining the CNOT [SWAP] from any operation of 
the corresponding class and provide a protocol to achieve this task.
\end{abstract}


\narrowtext

Much of the attention in Quantum Information Theory [QIT] was focused in recent 
years on obtaining a qualitative and quantitative description of the 
entanglement properties of pure and mixed states.  Apart from the important link 
of this problem to the basic principles of QIT and Quantum Mechanics, a proper 
understanding of entanglement is also expected to lead to new applications in 
quantum communication and quantum computation. As entanglement ---mainly in the 
form of a maximally entangled state--- is a key ingredient for many applications 
in QIT, it became a focus of interest to consider entanglement as a valuable resource. In 
this context, it is important to know how to manipulate entangled states in an 
optimal way,  because this determines which QIT tasks can be done (with optimal 
probability of success) using a given entangled state.

Quite recently, it was realized that entanglement of states is not the full story ---one may also consider the 
entanglement properties of  {\it quantum 
dynamics}. This is motivated in part by the facts that after all we deal with 
interactions in experiments, and the interactions allow us to create entangled 
states. This makes a proper understanding of qualitative and quantitative 
entanglement properties of quantum operations highly desirable. In fact, first 
steps in this direction have been reported recently 
\cite{Ci00,createentanglement,Kr00,createU,simham}.
However, even in the case of bipartite unitary dynamics ---which may in some 
sense be considered as the analog of bipartite pure states---, only very few is 
known up to now. Important issues, such as the structure of non--local unitary 
operations as well as the possible local interconvertability of non--local 
unitary operations ---i.e. the (probabilistic) simulation of an unitary 
operation using some other one--- are completely unexplored yet. Similarly as for  
pure state conversion, this last issue provides information about which kind of 
{\it dynamics} may be performed (with an optimal probability of success) given 
some unitary operation and allows us to put some structure in the set of non--local operations. 

Apart from the theoretical interest of such questions, they may also be of 
practical relevance. In particular, in many cases it is important to know 
whether a given resource (i.e. a certain unitary operation) suffices to 
implement a relevant task in QIT. For example, one may consider the the process 
of entanglement purification \cite{Be96}, one of the basic primitives for long 
range quantum communication using quantum repeaters \cite{Br98}. The known 
schemes for entanglement purification \cite{Be96} require the possibility to 
implement CNOT operations \cite{noteCNOT} between two particles. The approach 
sketched above allows to decide whether a given unitary operation ---e.g. 
produced by a weak interaction--- is already sufficient to implement a CNOT 
operation and thus entanglement purification. As entanglement purification 
itself is already a probabilistic process, one can consider probabilistic 
simulation of the operations. In this letter, we provide a complete solution of 
the problem stated above. From our results follows e.g. that any (arbitrarily 
weak) bipartite unitary operation already allows to implement entanglement 
purification.

We consider two (spatially) separated parties and non--local 
unitary operations $U, \tilde U$ acting on two $d$--level systems, each of them 
hold by one of the parties. We say that an unitary operation $U$ can simulate 
the action of another operation $\tilde U$ on an arbitrary input state 
probabilistically (equivalently,  $U$ can be converted to $\tilde U$), $U 
\rightharpoonup \tilde U$, if there exists a protocol using (stochastic) local 
operations and classical communication [(S)LOCC] \cite{Be99} and a {\it single} 
application of $U$ to achieve this task. The most general simulation protocol in 
this context consists of appending local auxiliary systems, a sequence of 
(S)LOCC, followed by a single application of $U$ and another sequence of 
(S)LOCC. 
We have to know when the protocol succeeds. In this case ---which 
occurs with some non--zero probability of success--, the total action of the 
protocol on an arbitrary input state $\rho$ ---after tracing out eventually 
involved auxiliary systems--- is given by $\tilde U \rho \tilde U^\dagger$.
Note that we do not allow for manipulations during the evolution (as is 
e.g. done in Hamiltonian simulation \cite{simham}). Such a situation appears if 
the process of interaction is inaccessible for some reason, e.g. because the 
interaction is taking place at a very short timescale, or the unitary operation 
is given by a fixed device (black box), which should then be used for some other 
purpose. 

One may also consider simulation of unitaries under other restricted 
sets of local operations, such as local unitaries (LU) or local operations 
without classical communication. For each kind $X$ of local operations, one 
obtains in this way an equivalence relation based on interconversion of $U$ and 
$\tilde U$, $U \rightleftharpoons_{X}  \tilde U$,  which is discussed in detail 
in \cite{Du01CU}. Throughout this letter, we will consider probabilistic 
interconversion and allow for arbitrary local resources \cite{notelocal}, i.e. 
SLOCC. 

The results we obtain are as follows: 
 \begin{itemize}
\item[(i)] We provide a necessary and sufficient 
condition for probabilistic simulation of unitary operations $U 
\rightharpoonup \tilde U$. This allows us to obtain a complete, hierarchic classification based on the 
equivalence relation induced by probabilistic interconvertability 
of operations, i.e. $U \rightleftharpoons \tilde U$. 
The number of classes is 
finite and for a $\C^d\otimes\C^d$ system, i.e. $U\in SU(d^2)$, bounded from above by $d^2$. 

\item[(ii)] For qubits, 
i.e. $d=2$, we show that only three classes exist, which can be represented by the 
identity, CNOT and SWAP respectively \cite{noteCNOT}.  

\item[(iii)] Also for the case of qubits, we explore the internal structure of the classes and show that given 
the CNOT [SWAP] operation, one can {\it deterministically} simulate any operation 
of its class. We also calculate the {\it optimal probability} to obtain 
the CNOT [SWAP] operation from an arbitrary operation of the corresponding class 
and provide a practical protocol to achieve this task.  
\end{itemize}
In order to obtain a necessary and sufficient condition for 
probabilistic gate simulation (i), we make use of the isomorphism 
between non--local physical operations and states \cite{Ci00}. When applied to 
unitary operations, one finds that to each unitary operation $U\in SU(d^2)$ 
corresponds a pure state $|\Psi_U\rangle \in \C^{d^2}\otimes\C^{d^2}$ 
given by 
\be
|\Psi_U\rangle\equiv U_{A_1B_1} |\Phi\rangle_{A_1A_2}\otimes|\Phi\rangle_{B_1B_2},\label{psiU}
\ee
where $|\Phi\rangle_{A_1A_2}\equiv1/\sqrt{d}\sum_{k=1}^d|k\rangle_{A_1}|k\rangle_{A_2}$ is a (local) maximally 
entangled state. The isomorphism has a very simple interpretation 
\cite{Ci00}: (a) On the one hand, Eq. (\ref{psiU}) tells us that one can 
obtain the state $|\Psi_U\rangle$ starting from a product state (systems A - B) 
given a single application of $U$. (b) On the other hand ---as shown in 
\cite{Ci00}--- given the state $|\Psi_U\rangle$, one can {\it 
probabilistically} implement the unitary operation $U$ on an arbitrary input 
state $\rho$ by performing suitable local measurements. We denote by 
$n_{\Psi_U}$ the Schmidt number of the state $|\Psi_U\rangle$, i.e. the number of 
non--zero Schmidt coefficients. Recall that (c) a bipartite pure state 
$|\psi\rangle$ can be transformed into another pure state $|\phi\rangle$ with 
non zero probability of success using SLOCC iff $n_\psi \geq n_\phi$ (see e.g. \cite{LoPo}). 

We can now state the following necessary and sufficient condition for probabilistic gate simulation:
\be
U \rightharpoonup \tilde U ~{\rm iff}~ n_{\Psi_U} \geq  n_{\Psi_{\tilde U}} \label{Usim}.
\ee
Using a sequence of local operations and the properties (a-c), the proof of Eq. 
(\ref{Usim}) can be summarized as follows: Necessity follows from the existence 
of the (probabilistic) local process $|\Psi_U\rangle \stackrel{(b)}{\rightarrow}  
U \rightarrow  \tilde U \stackrel{(a)}{\rightarrow} |\Psi_{\tilde U}\rangle$, 
which according to (c) implies that $n_{\Psi_U} \geq  n_{\Psi_{\tilde U}}$. 
Regarding sufficiency, we have: $U \stackrel{(a)}{\rightarrow} |\Psi_U\rangle 
\stackrel{(c)}{\rightarrow}  |\Psi_{\tilde U}\rangle 
\stackrel{(b)}{\rightarrow}\tilde U$. Note that the last relation provides a 
protocol to achieve the simulation $U \rightharpoonup \tilde U$.  

It immediately follows that $U$ can be interconverted into $\tilde U$ 
probabilistically, $U \rightleftharpoons \tilde U$, iff $n_{\Psi_U} =  
n_{\Psi_{\tilde U}}$. We have that two operations $U, \tilde U$ belong to the 
same equivalence class induced by this equivalence relation iff their corresponding pure states 
$|\Psi_U\rangle|, |\Psi_{\tilde U}\rangle$ have the same Schmidt number. Since 
$n_{\Psi_U} \leq d^2$, we have that at most $d^2$ inequivalent classes exist if 
$U\in SU(d^2)$. The classification is hierarchic, as unitary operations 
corresponding to states with a higher Schmidt number can simulate operations 
with corresponding states with a lower Schmidt number.

In order to make this more explicit, we will now turn into the probabilistic 
conversion of two--qubit gates, i.e. $d=2$ (ii). We  explicitly obtain the 
corresponding classes, which turn out to be only three. As shown by Kraus et. al. 
\cite{Kr00} (see also \cite{Kh01}), any two--qubit unitary operation $U\in SU(4)$ 
can be uniquely \cite{noteunique} written in the following standard form
 \bma\label{Ugen}\bea
U_{AB}&=&V_A\otimes W_B e^{-iH} \tilde V_A\otimes \tilde
W_B,\\
H&\equiv&\sum_{i=1}^3H_i, ~~~H_i\equiv\mu_i \sigma_i^A\otimes \sigma_i^B, \\
&&\pi/4\geq \mu_1\geq\mu_2 \geq |\mu_3| \geq 0,
\eea\ema
where $V_A,W_B,\tilde V_A,\tilde V_B$ are local unitary operations. That is, up 
to local unitaries, any two--qubit unitary operation is given by the normal form 
$e^{-iH}$, which contains all the non--local content of the operation. We denote 
by $\{|\Phi_i\rangle\}_{i=0,1,2,3}$ a basis of maximally entangled states, where  
$|\Phi_i\rangle\equiv\sigma_i\otimes \eins |\Phi\rangle$ and 
$|\Phi\rangle\equiv1/\sqrt{2}(|00\rangle+|11\rangle)$. We have that 
$|\Psi_U\rangle=V_{A_1}\otimes W_{B_1}\otimes \tilde V^T_{A_2} \otimes \tilde 
W^T_{B_2}e^{-iH_{A_1B_1}}|\Phi\rangle_{A_{1,2}}|\Phi\rangle_{B_{1,2}}=_{\rm LU} 
\sum_{k=0}^{3} a_i |\Phi_i\rangle_{A_{1,2}}|\Phi_i\rangle_{B_{1,2}}$. We used 
that $\eins_{A_1}\otimes V_{A_2} |\Phi\rangle = V^T_{A_1}\otimes \eins_{A_2} 
|\Phi\rangle$ in the first equality, while the second equality is understood up 
to local unitaries and corresponds ---up to some irrelevant phase factors--- to 
the Schmidt decomposition of $|\Psi_U\rangle$ with corresponding Schmidt 
coefficients $|a_i|$. We also used that $e^{-iH}=\sum_{k=0}^{3}a_i 
\sigma_i^A\otimes\sigma_i^B$. The coefficients $a_i$ are given by    
\bma\label{ai}\bea
a_0&=&c_{\mu_1}c_{\mu_2}c_{\mu_3}-is_{\mu_1}s_{\mu_2}s_{\mu_3},\\ 
 a_1&=&c_{\mu_1}s_{\mu_2}s_{\mu_3}-is_{\mu_1}c_{\mu_2}c_{\mu_3},\\ 
 a_2&=&s_{\mu_1}c_{\mu_2}s_{\mu_3}-ic_{\mu_1}s_{\mu_2}c_{\mu_3},\\ 
 a_3&=&s_{\mu_1}s_{\mu_2}c_{\mu_3}-ic_{\mu_1}c_{\mu_2}s_{\mu_3}, 
\eea\ema
and we introduced the shorthand notation $c_{\mu_i}\equiv \cos(\mu_i)$, 
$s_{\mu_i} \equiv \sin(\mu_i)$. 
It is now straightforward to check that the 
Schmidt number $n_{\Psi_U}$ 
is either 1,2 or 4.  Thus three classes of two--qubit unitary operations under probabilistic 
local interconversion, $U \rightleftharpoons \tilde U$, exist:
\\
{\bf Class 1: $n_{\Psi_U}=1:$} These are {\it local} unitary
  operations, with $\mu_1=\mu_2=\mu_3=0$ in Eq. (\ref{Ugen}). One can choose the identity as a representative of this class.
\\  
{\bf Class 2: $n_{\Psi_U}=2:$} 
These are non--local unitary operations with $\mu_1\not=0$ and $\mu_2=\mu_3=0$ in 
Eq. (\ref{Ugen}). The CNOT operation appears as a natural representative of 
this class, as $|\Psi_{U_{\rm 
CNOT}}\rangle=\frac{1}{\sqrt{2}}(|00\rangle_{A_{1,2}}|\Phi_0\rangle_{B_{1,2}}+|1 
1\rangle_{A_{1,2}}|\Phi_1\rangle_{B_{1,2}})$ is a maximally entangled state with 
Schmidt number 2. Note that the CNOT is up to local unitaries equivalent to an 
operation of the form (\ref{Ugen}) with $\mu_1=\pi/4,\mu_2=\mu_3=0$. 
\\
{\bf Class 3: $n_{\Psi_U}=4:$} 
These are non--local unitary operations  with $\mu_1, \mu_2\not=0$ and arbitrary 
$\mu_3$. The SWAP operation appears as a natural representative of this class, 
as $|\Psi_{U_{\rm SWAP}}\rangle=|\Phi_0\rangle_{A_1B_2}|\Phi_0\rangle_{A_2B_1}$ 
is a maximally entangled state with Schmidt number 4. Note that the SWAP is up 
to local unitaries equivalent to an operation of the form (\ref{Ugen}) with 
$\mu_1=\mu_2=\mu_3=\pi/4$. 

Recall that any operations of class 3 can simulate operations of class
2 probabilistically, however the reverse process is not possible. This
implies on the one hand that any non--local unitary operation can be
used to simulate a CNOT operation probabilistically (and thus to
implement entanglement purification), while the CNOT operation can
e.g. not be used to simulate $e^{-it(\sigma_1^A\otimes\sigma_1^B +
  \sigma_2^A\otimes\sigma_2^B)}$ with non--zero probability of success
even for $t\ll 1$.

We now explore the internal structure of the different classes (iii). 
We first show that the representative of each class can be used to simulate any other 
operation within the class with unit probability. For the SWAP operation, this 
is trivial to see: Since the SWAP operation can be used to create two ebits of 
entanglement (see Eq. (\ref{psiU})), one can use the first ebit to teleport a 
qubit from $B$ to $A$, implement the desired operation locally in $A$ and 
teleport the qubit back to $B$. Note that there is a whole set of operations 
within class 3 which can be deterministically interconverted into 
the SWAP operation. In particular, all operations of the form (\ref{Ugen}) with 
$\mu_1=\mu_2=\pi/4$ and arbitrary $\mu_3$ can create two ebits of entanglement 
and can thus similarly be used to deterministically implement any other operation.
Note that these are the only two-qubit operations with this property. 

We now show that the CNOT can simulate any operation of the form 
$U(\alpha)\equiv e^{-i\alpha \sigma_3^A\otimes\sigma_3^B}$ with $0\leq\alpha\leq\pi/4$ 
deterministically. This is sufficient to obtain that any operation in class 2 can 
be simulated by the CNOT, as any operation in class 2 is equivalent to 
$U(\alpha)$ up to LU. Given an arbitrary input state $|\phi\rangle_{AB}=\sum_{i,j} 
b_{ij}|i\rangle_A|j\rangle_B$ , we use the following protocol: (a) We apply 
CNOT$_{A\rightarrow \tilde B}$, where $\tilde B$ is an auxiliary system at site 
$B$ prepared in state $|0\rangle$.  (b) We apply locally the operation 
$U(\alpha)_{B \tilde B}$. (c) We measure the particle $\tilde B$ in the x-basis, 
i.e. we measure projectors corresponding to the states $|\pm 
\rangle\equiv 1/\sqrt{2}(|0\rangle \pm |1\rangle)$. If we obtain the result 
corresponding to $|-\rangle$, we apply $\sigma_3$ in A and $\tilde B$, otherwise 
we do nothing (the last step requires classical communication). It is 
straightforward to check that the protocol performs the desired gate on an 
arbitrary input state: $\sum_{i,j} b_{ij}|i\rangle_A|j\rangle_B 
|0\rangle_{\tilde B} \stackrel{(a)}{\rightarrow}  \sum_{i,j} b_{ij}|i\rangle_A|j\rangle_B 
|i\rangle_{\tilde B} \stackrel{(b)}{\rightarrow}  \sum_{i,j} (c_\alpha -i 
s_\alpha(-1)^{\delta_{j1}+\delta_{i1}}) b_{ij}|i\rangle_A|j\rangle_B 
|i\rangle_{\tilde B} \stackrel{(c)}{\rightarrow}  \sum_{i,j} (c_\alpha -i 
s_\alpha(-1)^{\delta_{j1}+\delta_{i1}}) b_{ij}|i\rangle_A|j\rangle_B 
|+\rangle_{\tilde B} = U_{AB}|\phi\rangle_{AB}|+\rangle_{\tilde B}.$

In the remainder of this paper, we will consider {\it optimal} simulation of 
CNOT [SWAP] given an arbitrary operation within the corresponding class, that is 
we are looking for the simulation protocol with highest possible probability of 
success. We show the following: (a) If a protocol can probabilistically simulate 
the action of a unitary operation on {\it all} input states, then the 
probability of success is {\it independent} of the input state. (b) We derive an 
upper bound for the probability of success for a particular input state, which 
---according to (a)--- provides also a bound for the optimal protocol. (c) We 
provide a protocol which reaches the upper bound. 

Regarding (a), we use that any simulation protocol can be considered as a 
two--branch protocol with two possible outcomes: The successful branch described 
by the operator $M$, where with some probability of success $U$ is performed 
and the second branch where the simulation of $U$ failed. We consider the 
successful branch and assume that the probability of success, $p_\psi={\rm 
tr}(M|\psi\rangle\langle\psi|M^\dagger)$ of the protocol depends on the input 
state $|\psi\rangle$, where $|\psi\rangle\equiv\sum_{ij}\alpha_{ij}|ij\rangle$. 
We have that $M|\psi\rangle = \sqrt{p_\psi} U|\psi\rangle$.  Using that 
$M|ij\rangle= \sqrt{p_{ij}}U|ij\rangle$, we have on the one hand 
$M\sum_{ij}\alpha_{ij}|ij\rangle = \sqrt{p_\psi}U 
\sum_{ij}\alpha_{ij}|ij\rangle$ and on the other hand 
$M\sum_{ij}\alpha_{ij}|ij\rangle=\sum_{ij}\alpha_{ij}\sqrt{p_{ij}}U|ij\rangle$. 
This implies that $\sqrt{p_\psi}=\sqrt{p_{ij}} ~\forall ij$ and thus 
$p_\psi\equiv p$ is constant.

(b) It follows from (a) that we can consider an arbitrary input state to derive 
an upper bound for the probability of success of the optimal protocol. As CNOT 
[SWAP] can create one [two] ebit(s) of entanglement when applied to product 
input states  (see Eq. (\ref{psiU})), the optimal simulation  protocol ---when 
acting on a product input state--- should also be capable of creating one [two] 
ebit(s) of entanglement. 
The first sequence of SLOCC does not 
change the product structure of the input state. Thus we have to consider the optimal 
probability to obtain a maximally entangled state with one [two] ebit(s) of 
entanglement using a single application of $U$ acting on an arbitrary product 
input state followed by SLOCC. Let $|\phi\rangle_A|\chi\rangle_B$ be 
an arbitrary product state, already including local auxiliary particles. We 
denote 
$|\phi_i\rangle_{A_1A_2}\equiv\sigma_i^{A_1}\otimes\eins_{A_2}|\phi\rangle_{A_1A
_2}$ and 
$|\chi_i\rangle_{B_1B_2}\equiv\sigma_i^{B_1}\otimes\eins_{B_2}|\chi\rangle_{B_1B
_2}$, which are normalized, but not necessarily orthogonal states. Using the 
notation introduced between Eq. (\ref{Ugen}) and Eq. (\ref{ai}) and assuming 
without loss of generality that $U=e^{-iH}$ (the local unitary operations of a 
general $U$ of Eq. (\ref{Ugen}) can be absorbed into the definition of 
$|\phi\rangle,|\chi\rangle$ and the SLOCC performed afterwards), we 
obtain $|\chi_U\rangle \equiv U_{A_1B_1}|\phi\rangle_A|\chi\rangle_B = 
\sum_{k=1}^3 a_k |\phi_k\rangle_A|\chi_k\rangle_B$. Let $b_i$ be the ordered 
Schmidt coefficients of $|\chi_U\rangle$, $b_0\geq b_1\geq b_2\geq b_3$. Using 
the results of \cite{LoPo,Vixx} on optimal conversion of pure states, we have 
that the optimal probability to obtain a maximally entangled state of one [two] 
ebit(s) out of $|\chi_U\rangle$ is given by $p_1=2(b_1^2+b_2^2+b_3^2) 
[p_2=4b_3^2]$. Note that for operations in class 2, the maximal probability 
$p_1$ reduces to $p_1=2b_1^2$, since $b_2=b_3=0$ in this case. Let $|\tilde 
\phi_j\rangle_{A}$ be a normalized state for which $|\langle \tilde\phi_j|\phi_i\rangle|=0$ 
for $i\not=j$.  For $|\tau\rangle_{A}$ being an arbitrary state and 
$\rho_{A}\equiv {\rm tr}_{B_1B_2}(|\chi_U\rangle\langle\chi_U|)$ the reduced 
density matrix of system $A$, we have that $b_3^2\leq \langle\tau_A|\rho_A|\tau_A 
\rangle$. For $|\tau_A\rangle=|\tilde \phi_3\rangle$, we obtain $b_3^2\leq 
\langle\tilde\phi_3|\rho_A|\tilde\phi_3\rangle \leq |a_3|^2 
|\langle\tilde\phi_3|\phi_3\rangle|^2 \leq  |a_3|^2$. This implies that $p_2 
\leq 4|a_3|^2$. A similar argument can be used to obtain a bound for $p_1$ for 
unitary operations within class 2, however one has to take into account that 
$b_2=b_3=0$ and thus obtains $b_1^2 \leq |a_1|^2$, which implies $p_1 \leq 
2|a_1|^2=2s_{\mu_1}^2$.

(c) The optimal protocol 
is given by the one sketched in the proof of Eq. (\ref{Usim}), i.e. $U 
\rightarrow |\Psi_U\rangle \rightarrow  |\Psi_{U_{CNOT [SWAP]}}\rangle 
\rightarrow \tilde U_{CNOT} [U_{SWAP}]$, where in the second step the optimal 
protocol for pure state conversion \cite{LoPo,Vixx} is used. Note that 
the first and third step can be performed with unit probability 
of success, where the last relation follows from the fact that one [two] ebit(s) 
of entanglement are sufficient to implement a CNOT [SWAP] deterministically 
\cite{Ci00,createU}. We have that the conversion probability to obtain a CNOT 
[SWAP] given an operation in class 2 [3] is $p_1=2|a_1|^2=2s_{\mu_1}^2$ 
[$p_2=4|a_3|^2$], which reaches the upper bound derived above and the protocol 
is thus optimal.    

In this letter, we derived necessary and sufficient conditions for probabilistic 
(inter)conversion of bipartite unitary operation and obtained a complete, 
hierarchic classification. For two--qubit operations, we proved the existence 
of three inequivalent classes, represented by $\eins_{AB}$, CNOT and SWAP. We 
provided protocols to obtain the {\it optimal} conversion between the 
representative of each class and an arbitrary operation of the class.  Note that 
the results presented in this letter can be extended to multipartite systems 
and one may also consider interconversion of operations under restricted classes of local operations 
\cite{Du01CU}.


This work was supported by European Community under project EQUIP (contract 
IST-1999-11053), grant HPMF-CT-2001-01209 (W.D.) and grant HPMF-CT-1999-00200 
(G.V.) (Marie Curie fellowships),  the ESF and 
the Institute for Quantum Information GmbH . 



\begin{references}
\bibitem{noteCNOT}
The CNOT [SWAP] operation is defined by the following mapping of states, written in the
standard basis: $|i\rangle_A|j\rangle_B \rightarrow |i\rangle_A|i\oplus
j\rangle_B$ [$|i\rangle_A|j\rangle_B \rightarrow |j\rangle_A|i\rangle_B$], where $\oplus$ denotes addition modulo 2.
\end{references}
\end{document}